\documentstyle[prd,tighten,aps]{revtex}
\begin{document}
\draft

\title{\bf Anti-de Sitter wormhole kink}
\author{Pedro F. Gonz\'alez-D\'{\i}az}
\address{Centro de F\'{\i}sica ``Miguel Catal\'an'',
Instituto de Matem\'aticas y F\'{\i}sica Fundamental,\\
Consejo Superior de Investigaciones Cient\'{\i}ficas,
Serrano 121, 28006 Madrid (SPAIN)}
\date{May 19, 1997}

\maketitle

\begin{abstract}

The metric describing a given finite sector of a four-dimensional
asymptotically anti-de Sitter wormhole can
be transformed into the metric of the time constant sections
of a Tangherlini black hole in a
five-dimensional anti-de Sitter spacetime when one allows light cones
to tip over on the hypersurfaces according to the conservation laws
of an one-kink. The resulting kinked metric can be maximally extended,
giving then rise to an instantonic structure on the euclidean continuation
of both the Tangherlini time and the radial coordinate. In the
semiclassical regime, this kink is related to the existence of
closed timelike curves.

\end{abstract}

\renewcommand{\theequation}{\arabic{section}.\arabic{equation}}

\section{\bf Introduction}
\setcounter{equation}{0}

The choice of the boundary conditions for the quantum state of the
universe should be carried out on just the two natural possibilities
which exist for positive definite metrics; i.e.: compact metrics or
noncompact metrics which are asymptotic to metrics of maximal
symmetry. Choosing either only compact metrics to avoid$^{1}$ any
boundary for the quantum state of the universe, or noncompact
metrics which are disconnected, consist of a compact part with
physical boundary at a given hypersurface and an asymptotically
euclidean or anti-de Sitter part without any inner boundary, and
dominate on any noncompact connected metric$^{2,3}$, does not make
much difference for the resulting state amplitude. For practical
purposes, instead of compact metrics$^{1}$, one could likewise take
noncompact disconnected metrics for the quantum state of the
universe. Therefore, metrics which are asymptotically euclidean
or anti-de Sitter may also contribute the path integral describing
the state of the universe. Clearly, when such metrics are endowed
with a given nonzero boundary which behaves as a microscopic
bridge between two asymptotic, large regions, these metrics
would represent contributions from wormholes to the quantum
state of the universe.

Wormholes whose maximally-symmetric asymptotic metrics are flat
space have already been extensively considered$^{4}$. They come
about as solutions of the euclidean Einstein equations for gravity
coupled to special kinds of matter fields$^{5}$. Contributing
connections bridging maximally-symmetric anti-de Sitter
asymptotic metrics were first studied in Refs. 6 and 7 for the
case of a massless, scalar field which conformally couples to
gravity.

Wormhole kinks$^{8}$ can be regarded as topologically admissible
generalizations of euclidean wormholes, allowing for a connection
between topology changes and black holes$^{9}$.
Whereas asymptotically euclidean wormholes and their
kinky extensions$^{8}$ have also been studied in terms of
spherically symmetric coordinates in spacetimes that show an
apparent singularity at the throat radius$^{9}$, the corresponding
spherically symmetric representation and kinky extension of
wormholes which are asymptotically anti-de Sitter have not
hitherto been considered. Since these wormholes have vanishing
action (as it can be seen$^{7}$ by addition of suitable surface
terms and gauge fixing) one should expect that they would also
naturally contribute the probability amplitude for the universe.

The present work aims at bringing
such a study to completion by regarding the wormhole metric
as a constant-time section of a five-dimensional black hole
in anti-de Sitter space, and discussing the possible implications
that the existence of asymptotically anti-de Sitter wormhole
kinks may have in the presence of closed timelike curves.

Let us first briefly review the general topological concept
of a kink and its associated topological charge.
Let $({\bf M},g_{ab})$ be a given D-dimensional spacetime, with
$g_{ab}$ a Lorentz metric on it. One can always regard $g_{ab}$
as a map from any connected D-1 submanifold $\Sigma\subset {\bf M}$
into a set of timelike directions in ${\bf M}$ $^{10}$. Metric homotopy
can then be classified by the degree of this map. This is
seen by introducing a unit line field $\{n,-n\}$, normal to
$\Sigma$, and a global framing $u_{i}$: $i$=1,2,...,D-1, of
$\Sigma$. A timelike vector ${\bf v}$ can then be written in terms
of the resulting tetrad framing $(n,u_{i})$ as
$v=v^{0}n+v^{i}u_{i}$, such that $\sum_{i}^{D-1}(v^{i})^{2}=1$.
Restricting to time orientable manifolds ${\bf M}$, ${\bf v}$ then
determines a map
\[K: \Sigma\rightarrow S^{D-1}\]
by assigning to each point of $\Sigma$ the direction that
${\bf v}$ points to at that point. This mapping allows a general
definition of kink and kink number. Respect to hypersurface
$\Sigma$, the kink number (or topological charge) of the
Lorentz metric $g_{ab}$ is defined by$^{10}$
\[{\rm kink}(\Sigma;g_{ab})={\rm deg}(K),\]
so this topological charge measures$^{11}$ how many times the
light cones rotate all the way around as one moves
along $\Sigma$.

In the case of an asymptotically flat spacetime the pair
$(\Sigma,g)$ will describe an asymptotically flat kink if
${\rm kink}(\Sigma;g)\neq 0$. All of the topological charge of
the kink in the metric $g$ is in this case confined to
some finite compact region$^{11}$. Outside that region all
hypersurfaces $\Sigma$ are everywhere spacelike.
Topology changes, such as handles or wormholes, can occur
in the compact region supporting the kink, but not outside
it. All topologies are actually allowed to happen in such
a region. Therefore, in the case of spherically symmetric
kinks, the supporting region should be viewed as an
essentially quantum-spacetime construct. This is the
view we shall assume throughout this paper.

Imposing to the wormhole space the invariance properties
of an one-kink would induce a change in the geometrical
structure of the wormhole that resulted in a bridge with
smaller cross-sectional area at the neck$^{9}$. When visualized as
a constant-time section of a five-dimensional black hole in
anti-de Sitter space, one would moreover expect the wormhole
metric to correspond to some four-dimensional Einstein-Rosen
bridge$^{12}$ in anti-de Sitter space. This space has two
problems. First, on the maximal analytical extension of
its metric, one of the two wormhole halves must necessarily
be described in the nonphysical exterior region$^{13}$, and
secondly, this wormhole should inexorably pinch off at the
neck, just as it happens in the three-dimensional Einstein-Rosen
bridge$^{12,13}$.
These difficulties would no longer be expected to occur in
the corresponding kinked wormhole. Since the kink number must
be conserved, the wormhole ought to be described in two distinct
coordinate patches only identified at a given common surface
inside the compact region supporting the kink, with a black
hole living in each patch. This identification would occur
both on the physical and nonphysical regions of the maximal
analytical extension, separately, and would represent a bridge
which could not pinch off.

The paper is organized as follows. In Sec. II we deal with the
metric of asymptotically anti-de Sitter wormholes in terms of
spherically symmetric coordinates and avoid its singularity
at the neck by a convenient coordinate transformation. Sec. III
generalizes this metric to the case where there exists an one-kink.
The geodesic incompleteness of the resulting metric is removed by
Kruskal extension in Sec. IV and in Sec. V we analytically
continue the Kruskal kink metric into the euclidean regime, discussing
the instantons that represent formation of black-hole kinks in
anti-de Sitter spacetime. We consider the thermal emission of
these black holes and the related possibility that the kink be
linked to the formation of closed timelike curves in Sec. VI.
Finally, we summarize and conclude in Sec. VII.

\section{\bf Spherically-symmetric wormholes in anti-de Sitter space}
\setcounter{equation}{0}

Wormholes in asymptotically anti-de Sitter spacetime were first
considered in$^{6}$ and dealt with in greater detail in$^{7}$. They
correspond to solutions of the euclidean Einstein equations for
Hilbert-Einstein gravity with a negative cosmological constant,
conformally coupled to a massless scalar field. The semiclassical
geometry of these wormholes is described by the isotropic and
homogeneous metric
\begin{equation}
ds^{2}=a(\eta)^{2}\left(d\eta^{2}+d\Omega_{3}^{2}\right),
\end{equation}
in which $d\Omega_{3}^{2}$ is the metric on the unit three-sphere,
the scale factor $a(\eta)$ is given by
\begin{equation}
a(\eta)=\left(\frac{b-1}{2\Lambda}\right)^{\frac{1}{2}}nc(b^{\frac{1}{2}}\eta),
\end{equation}
with $b=(1+4\Lambda M^{2})^{\frac{1}{2}}$, $\Lambda$ is the absolute
value of three times the cosmological constant, $M$ an integration
constant, $nc$ the Jacobian elliptic function, and $\eta$ the
conformal time $\eta=\int\frac{d\tau}{a}$, where $\tau$ is the
Robertson-Walker time.

We note that metric (2.1), (2.2) can also be written in the static,
spherically-symmetric form
\begin{equation}
ds^{2}=\left(1-\frac{M^{2}}{r^{2}}+\Lambda r^{2}\right)^{-1}dr^{2}
+r^{2}d\Omega_{3}^{2},
\end{equation}
where $M$ can be taken to be the same mass constant as that
appearing in the definition of parameter $b$ above, and $\Lambda$
is renormalized so that it becomes also the same as the cosmological
constant in $b$.

While it is rather trivial to see that the Robertson-Walker metric
given by (2.1) and (2.2) describes the space of a wormhole with
a neck at $a(\eta=0)$, it is not so obvious to show that the
spherically symmetric metric (2.3) describes a wormhole. In
particular, inspection of (2.3) tells us that there is a
singularity at $r=\left(\frac{b-1}{2\Lambda}\right)^{\frac{1}{2}}$
and, therefore, metric (2.3) is geodesically incomplete. In what
follows, we shall extend this metric so that it, in fact, becomes
regular, with a neck replacing the singularity.

The singularity at $r=\left(\frac{b-1}{2\Lambda}\right)^{\frac{1}{2}}$
corresponds to an apparent coordinate singularity, rather than a true
curvature singularity. It can be avoided by introducing a new coordinate
from the definition
\begin{equation}
\frac{dr}{\left(1-\frac{M^2}{r^2}+\Lambda r^2\right)^{\frac{1}{2}}}
=rd\left[\ln u(r)\right],
\end{equation}
i.e:
\begin{equation}
\ln\left(\frac{u(r)}{\mu}\right)=
\frac{2i}{\sqrt{b\Lambda}}F\left[\arcsin\left(\frac{1-\sqrt{\frac{2\Lambda}{b-1}}r}{2}\right),\sqrt{b+1}\right],
\end{equation}
in which $\mu$ is an arbitrary integration constant and $F$ is the
elliptic integral of the first kind$^{15}$.

From (2.5) we finally obtain the coordinate transformation
\begin{equation}
r=\sqrt{\frac{2(b-1)}{\Lambda}}\left(nc\left[\frac{\sqrt{b\Lambda}}{2}\ln\left(\frac{u}{\mu}\right)\right]-\frac{1}{2}\right).
\end{equation}

Using (2.6) we transform (2.1) and (2.2) into the asymptotically
anti-de Sitter wormhole metric
\begin{equation}
ds^{2}=\left(\frac{2(b-1)}{\Lambda}\right) \left(nc\left[\frac{\sqrt{b\Lambda}}{2}\ln\left(\frac{u}{\mu}\right)\right]-\frac{1}{2}\right)^{2}
\left(\frac{du^2}{u^2}+d\Omega_{3}^{2}\right),
\end{equation}
which is everywhere regular and shows a neck at the arbitrary
scale $u=\mu$, i.e.: at $r=\sqrt{\frac{b-1}{2\Lambda}}$, as for metric
defined by (2.1) and (2.2). The
metric (2.7) gives the geometry of a regular spherically-symmetric wormhole
in an asymptotically anti-de Sitter spacetime.

\section{\bf The Tangherlini-anti-de Sitter kink}
\setcounter{equation}{0}

A worth-exploiting feature of metric (2.3) is that it corresponds
to the metric of a five-dimensional Tangherlini black hole in
anti-de Sitter space
\begin{equation}
ds^{2}=-\left(1-\frac{M^{2}}{r^{2}}+\Lambda r^{2}\right)dT^{2}+\left(1-\frac{M^{2}}{r^{2}}+\Lambda r^{2}\right)^{-1}dr^{2}
+r^{2}d\Omega_{3}^{2}
\end{equation}
at a surface with constant time $T=t_{0}$. The relation between
(2.3) and (3.1) can be viewed within the context of the kink
concept$^{8}$ in the following way:
the kink extension of metric (3.1) would also be the
kink extension of metric (2.3), with $T=t_0$,
corresponding to all possible
light cone orientations which are compatible with the existence
of one kink, starting with a future-directed light cone
orientation where the kinked metric should reduce to (2.3) [9].
Thus, we will consider metric (2.3)
to correspond to a given fixed value of time $T$ in the kink
extension of the Tangherlini anti-de Sitter metric, or in
other words, we wish to find an isometric embedding of the
four-dimensional wormhole described by metric (2.3), as
removed from (3.1), in a space which is the kinked extension
of (3.1).

We shall take for the metric that describes a three-spherically
symmetric general kink$^{9}$
\begin{equation}
ds^{2}=-\cos 2\alpha\left(dt^2-dr^2\right)-2\sin 2\alpha dtdr+r^2d\Omega_{3}^{2},
\end{equation}
where $\alpha$ is the angle of tilt for the light cones. An one-kink
is ensured to exist if $\alpha$ is allowed to monotonously increase from
0 to $\pi$, starting with $\alpha(0)=0$. Then, it can be checked
that metric (3.2) converts
into (2.3) if we use the substitution
\begin{equation}
\sin\alpha=\frac{1}{\sqrt{2}}\left(\frac{M^2}{r^2}-\Lambda r^2\right)^{\frac{1}{2}},
\end{equation}
with $0\leq\alpha\leq\pi$,
and introduce the change of time variable $t+g(r)=t_{0}$, where
$t_{0}$ is the arbitrary constant time labeling the surface on which
metric (2.3) is defined from (3.1), with the function $g(r)$ chosen such that
$\frac{dg}{dr}=\tan 2\alpha$.
%At first sight, the conversion of
%(3.2) into (2.3) using (3.3) and $t+g(r)=t_0$, could seem confusing
%since metrics (3.2) and (2.3) describe spaces of apparently differing
%dimensions. However, as $T=t_0$ is constant, time $t$ entering
%metric (3.2) is a function of $r$ alone
%and, therefore, the space
%described by (3.2) actually corresponds to the same number of
%dimensions as the space of metric (2.3). It follows, moreover,
%that metric is a solution of the same set of field equations
%as for (2.3) when the above change of variables is introduced
%in these equations.

For future-directed orientation of light
cones, $\alpha=0$, time $t$ becomes also constant and, in fact, metric
(3.2) reduces to (2.3), as it was pointed out at the beginning of
this section.

Since $\sin\alpha$ cannot exceed unity, it follows that
\[r_{max}=\left(\frac{M}{\sqrt{\Lambda}}\right)^{\frac{1}{2}}\geq r\geq
r_{min}=\left(\frac{a-1}{\Lambda}\right)^{\frac{1}{2}},\]
with
$a=\left(1+\Lambda M^2\right)^{\frac{1}{2}}$, so that $\alpha$
varies only from 0 to $\frac{\pi}{2}$. It is worth noticing
that the presence of a negative cosmological term in the metric
makes the asymptotic region $r\rightarrow\infty$ to vanish,
leaving a finite maximum surface at $r=r_{max}$, so that in this case
the kink turns out to be a gravitational topological defect with
finite volume.

A complete one-kink with finite volume
can only be achieved if we add a second
coordinate patch in order to describe the other half of the
$\alpha$-interval from $\frac{\pi}{2}$ to $\pi$. This extension
requires introducing a new time coordinate $\bar{t}=t+h(r)$,
where
\begin{equation}
h=\int dr\left(\frac{dg}{dr}-\frac{k}{\cos 2\alpha}\right),
\;\;\; k=\pm 1,
\end{equation}
which transforms (3.2) into the standard kinked metric$^{16}$
\begin{equation}
ds^2=-\cos 2\alpha d\bar{t}^2-2kd\bar{t}dr+r^2d\Omega_{3}^{2}.
\end{equation}
For $T=t_0$ constant, one can check that
time $\bar{t}$ is a function of $r$ alone, and that metric
(3.5) converts into metric (2.3) by using $\bar{t}=t_0-g(r)+h(r)$,
(3.3) and (3.4).

The zeros in the denominator of $\frac{dh}{dr}=
(\sin 2\alpha\mp 1)/\cos 2\alpha$ occur at the two horizons where
$r=r_h=\sqrt{\frac{b-1}{2\Lambda}}$, one per patch. For the first
patch, the horizon occurs at $\alpha=\alpha_h=\arcsin\left(\frac{b-a^2}{\sqrt{2}(b-1)}\right)$
and therefore the upper sign ($k=+1$) is selected so that $\frac{dh}{dr}$
and $h$ remain well defined and hence the kink is not lost in the
transformation from (3.2) to (3.5). The horizon in the second patch
takes place at $\alpha=\alpha_h+\frac{\pi}{2}$ and therefore the
lower sign ($k=-1$) is selected.

The need for two coordinate patches to describe the complete
spacetime of the kink can also be seen by regarding the
connection between the wormhole and the kink as an
embedding;
i.e. we visualize the wormhole metric (2.3) as removed from
Tangherlini spacetime and embedded in the kinked spacetime.
The embedding surface should then flare outward at $r=r_{max}$
and inward at $r=r_{min}$, regularly in both cases. The latter
requirement comes from the fact that the maximum cross section
of the wormhole is finite and its minimum cross section is
nonzero, i.e. the wormhole should avoid pinching off.

The metric (3.2) fails to be the metric of such an embedding;
for, though time $t$ is a function of $r$ interpretable as
an embedding function such that $dt/dr=-dg/dr=-\tan 2\alpha$,
and hence the rate of embedding
\[\frac{d^2 r}{dt^2}=\frac{\left(1-\frac{M^2}{r^2}+\Lambda r^2\right)
\left(\frac{M^2}{r^3}+\Lambda r\right)\left[2-\left(\frac{M^2}{r^2}-\Lambda r^2\right)^2\right]}
{\left(\frac{M^2}{r^2}-\Lambda r^2\right)^2\left(1-\frac{M^2}{2r^2}+\frac{\Lambda r^2}{2}\right)^2}\]
is positive (i.e. the embedding surface flares outward) at $r_{max}$
and negative (i.e. the embedding surface flares inward) at $r_{min}$,
$\frac{d^2 r}{dt^2}$ becomes singular at these extreme regions, so
that the wormhole pinches off at the second of them.

However, if we choose the standard kinked metric (3.5) and, therefore,
we take $\bar{t}=t_0-g+h$ as the embedding function, such that
$\frac{d\bar{t}}{dr}=-\frac{k}{\cos 2\alpha}$, then the rate of
embedding becomes independent of $k$ and given by
\[\frac{d^2 r}{d\bar{t}^2}=2\left(1-\frac{M^2}{r^2}+\Lambda r^2\right)
\left(\frac{M^2}{r^3}+\Lambda r\right),\]
which is positive at $r=r_{max}$,
\[\left.\frac{d^2 r}{d\bar{t}^2}\right|_{r_{max}}=\frac{4M^2}{r_{max}^{3}},\]
negative at $r_{min}$,
\[\left.\frac{d^2 r}{d\bar{t}^2}\right|_{r_{min}}=-4\frac{\left(\frac{M^2}{r_{min}^2}-1\right)}{r_{min}},\]
and regular everywhere. Thus, the wormhole turns out to be
connectible to the complete kinked spacetime with metric (3.5)
at both extreme surfaces of the wormhole; that is to say, the
wormhole (2.3) can be interpreted as a kinked boundary in the
space with metric (3.5), in each of the two coordinate patches.

Although still geodesically-incomplete, metric (3.5) will describe
the complete interval of the light-cone configurations running from
light cones that point vertically up and out from the finite
maximum surface at $\alpha=\pi$, to light cones that point
vertically up and into the other maximum surface at $\alpha=\pi$,
and this is only possible if two coordinate patches identified
on the minimum surface at $\alpha=\frac{\pi}{2}$ are used.

On the other hand, every kink is characterized by a topological
charge or kink number$^{9}$. If, as in the case being considered,
one has an one-kink, the topological charge can be either positive
when the light cones rotate away from external observers (that is
a black hole in anti-de Sitter space), or negative when the light
cones rotate in the opposite direction for the case of a white
hole in anti-de Sitter space. In the latter case, the sign of the
crossed metric component $g_{tr}$ in (3.2), or $g_{\bar{t}r}$ in
(3.5), is changed in each coordinate patch.

A specific characteristic of the considered kink is that, for
the real values of the angle $\alpha$ running from 0 to $\pi$,
the resulting kinked spacetime covers the sector of the
original kinkless metric going up to $r_{max}$ only. One still
could cover the spacetime up to $r=\infty$ by analytically
continuing the tilt angle so that $\alpha\rightarrow\alpha
+i\tilde{\alpha}$, allowing it to monotonously run first
from $\alpha=0$, $\tilde{\alpha}=+\infty$, at $r=\infty$, to
$\alpha=0$, $\tilde{\alpha}=0$, at $r_{max}$, along the first
corrdinate patch, then to $\alpha=\pi$, $\tilde{\alpha}=0$,
at $r_{max}$, after crossing to the second patch, and finally
to $\alpha=\pi$, $\tilde{\alpha}=-\infty$, at $r=\infty$, on
the second patch. Here, the rotation of the light cones over
the hypersurfaces is taken to occur in the complex $\alpha$-plane,
and the sectors of the original spacetime beyond $r_{max}$ are
covered by monotonous variation of the purely imaginary
parameter $\tilde{\alpha}$.

\section{\bf Kruskal extension}
\setcounter{equation}{0}

The standard kinked metric for black or white holes,
\[ds^2=-\cos 2\alpha d\bar{t}^2\mp 2kd\bar{t}dr+r^2d\Omega_{3}^{2},\]
still contains a geodesic incompleteness at the horizon
$r=r_h=\left(\frac{b-1}{2\Lambda}\right)^{\frac{1}{2}}$, in each patch.
These apparent singularities can be removed by using the Kruskal
technique$^{17}$. In order to achieve the maximally-extended metric,
let us introduce the general metric
\begin{equation}
ds^2=-H(U,V)dUdV+r^2d\Omega_3^2,
\end{equation}
where
\begin{equation}
H=\left(\frac{b-1}{2\Lambda}\right)^{\frac{1}{2}}\frac{\cos 2\alpha}{\beta}
\exp\left(-2k\beta\int_{r_{max}/r_{min}}^{r}\frac{dr}{\cos 2\alpha}\right)
\end{equation}
\begin{equation}
U=\mp e^{\beta\bar{t}}\exp\left(2k\beta\int_{r_{max}/r_{min}}^{r}\frac{dr}{\cos 2\alpha}\right)
\end{equation}
\begin{equation}
V=\mp\left(\frac{2\Lambda}{b-1}\right)^{\frac{1}{2}}\frac{e^{-\beta\bar{t}}}{2\beta};
\end{equation}
the constant $\beta$ in (4.2)-(4.4) is an adjustable parameter to be
chosen such that the unphysical singularity at $r=r_h$ is removed, and
the lower integration limit $r_{max}/r_{min}$ accounts for the choices
$r=r_{max}$ and $r=r_{min}$, depending on whether the first or second
patch is being considered. Using (3.3), we obtain for (4.2):
\[H=\frac{\sqrt{\Lambda(b-1)}\left(r^2-\frac{b-1}{2\Lambda}\right)\left(r^2+\frac{b+1}{2\Lambda}\right)}{\sqrt{2}\beta r^2}\]
\begin{equation}
\times\exp\left\{-\frac{\sqrt{2(b+1)}k\beta}{\sqrt{\Lambda}b}\arctan\left[\sqrt{\frac{2\Lambda}{b+1}}r\right]\right\}\times\left[\frac{\left(\sqrt{\frac{b-1}{2\Lambda}}+r\right)^2}{\frac{b-1}{2\Lambda}-r^2}\right]^{\sqrt{\frac{b-1}{2\Lambda}}\frac{
k\beta}{b}}.
\end{equation}

Eqn. (4.5) would actually have some constant term coming from the
lower integration limit $r_{max}/r_{min}$. This term has been omitted
because it is canceled by the similar constant term of the Kruskal
coordinate $U$ when forming the Kruskal metric.

Geodesic incompleteness is avoided if we choose
\begin{equation}
\beta=\frac{\sqrt{2\Lambda}b}{k\sqrt{b-1}}.
\end{equation}
Hence, we finally obtain for the Kruskal metric
\begin{equation}
ds^2=-\frac{k(b-1)\left(r^2+\frac{b-1}{2\Lambda}\right)\left[\sqrt{\frac{b-1}{2\Lambda}}+r\right]^{2}}{2br^2}
\times\exp\left[-2\sqrt{\frac{b+1}{b-1}}\arctan\left(\sqrt{\frac{2\Lambda}{b+1}}r\right)\right]dUdV+r^2d\Omega_3^2,
\end{equation}
with

\begin{equation}
U=\mp e^{bk\sqrt{\frac{2\Lambda}{b-1}}\bar{t}}\left(\frac{\sqrt{\frac{b-1}{2\Lambda}}-r}{\sqrt{\frac{b-1}{2\Lambda}}+r}\right)\times\exp\left[2\sqrt{\frac{b+1}{b-1}}\arctan\left(\sqrt{\frac{2\Lambda}{b+1}}r\right)\right]
\end{equation}
\begin{equation}
V=\mp\frac{ke^{-bk\sqrt{\frac{2\Lambda}{b-1}}\bar{t}}}{2b},
\end{equation}
where

\begin{equation}
\bar{t}=\bar{t}_{0}(k)-\frac{k}{b}\sqrt{\frac{b+1}{2\Lambda}}\arctan\left(\sqrt{\frac{2\Lambda}{b+1}}r\right)+\frac{k}{2b}\sqrt{\frac{b-1}{2\Lambda}}\ln\left(\frac{1+\sqrt{\frac{2\Lambda}{b-1}}r}{1-\sqrt{\frac{2\Lambda}{b-1}}r}\right);
\end{equation}
in (4.10) the constant coming from the lower integration limit has been
absorbed into the constant term $\bar{t}_{0}(k)$ whose value will therefore
depend on the coordinate patch we are considering.

By allowing $\alpha$ to be analytically continued to complex values,
and the lower integration limit for the first patch to become then
infinity, the integration in the expressions for $H$ and $U$ could
be performed along the continuous path described at the end of
Sec. IV, so obtaining a truly anti-de Sitter asymptotic behaviour
for the maximally extended kinked metric.
The resulting metric should be
compared with the extended asymptotically anti-de
Sitter wormhole metric (2.7) which is defined in one coordinate patch
only. It can be seen that the form of (4.7) coincides with that of
the Kruskal extension of the five-dimensional Tangherlini-anti-de
Sitter kink and differs from the kinkless counterpart of this just
by the presence in (4.7) of the sign parameter $k$. We also note that in the
limit $\Lambda\rightarrow 0$, (4.7) reduces to the kinked
five-dimensional black hole metric.

Continuity of the tilt angle also at $\frac{\pi}{2}$ ensures the
identification of the two coordinate patches on the surface $r=r_{min}$.
Such an identification should be done both on the original physical
regions and on the new, nonphysical regions created by the Kruskal
extension, separately. It gives rise to bridges that connect the
maximal surfaces at $r_{max}$ of the
asymptotically anti-de Sitter regions in the two patches. Any
constant time $T=\bar{t}_0$
sections of the resulting spacetime will then describe
four-dimensional wormholes occurring either in the physical or
nonphysical regions, whose neck is now at $r=r_{min}$, rather
than $r=\left(\frac{b-1}{2\Lambda}\right)^{\frac{1}{2}}$, in asymptotic
anti-de Sitter space, such as it was announced in the Introduction.

Finally, we can now see why the studied classical kink cannot be
linked to the presence of closed timelike curves. At first glance,
it might seem that the above discussed identification between the
two patches on $r=r_{min}$ would allow us to choose a null geodesic
such that it started in a given largest-surface of the
original regions and
would somehow "loop back" through the new regions to finally arrive
at its starting point. However, one can easily convince oneself
that such an itinerary is classically disallowed, since it requires
identification of the two patches also on maximum surfaces belonging
to a physical and a nonphysical region, respectively (see Fig. 1).

\section{\bf The instantons}
\setcounter{equation}{0}

Euclidean continuations of the metric that contains one kink must
be obtained by using the Wick rotation
\begin{equation}
\bar{t}\rightarrow i\bar{\tau},
\end{equation}
with $\bar{t}$ as given in (4.10). Because of time redefinition, the
instanton structure is richer than in the kinkless case. Thus, it
will be shown in what follows that the euclidean continuation (5.1)
should give rise to metrics which are positive definite only if we
choose either the usual continuation $\bar{t}_{0}\rightarrow i\bar{\tau}_{0}$,
for $r\geq r_h$ and $k=+1$, or the new kinky continuation $r\rightarrow
-i\rho$, $M\rightarrow -i\mu$, $\Lambda\rightarrow -\lambda$, for
$r<r_h$ and $k=-1$, where $r$ becomes timelike and we transform a
space coordinate into a time coordinate.

In order to see how this can be, let us consider the new variables$^{18}$
$y+z=U$ and $y-z=V$ in the Kruskal metric (4.7), which then becomes:
\begin{equation}
ds^2=-kG(\Lambda,M,r)(dy^2-dz^2)+r^2d\Omega_3^2,
\end{equation}
with

\begin{equation}
G(\Lambda,M,r)=\frac{k(b-1)\left(r^2+\frac{b-1}{2\Lambda}\right)\left[\sqrt{\frac{b-1}{2\Lambda}}+r\right]^{2}}{2br^2}\times\exp\left[-2\sqrt{\frac{b+1}{b-1}}\arctan\left(\sqrt{\frac{2\Lambda}{b+1}}r\right)\right].
\end{equation}

We have furthermore
\begin{equation}
y^2-z^2=kJ(\Lambda,M,r)\left(\frac{1-\sqrt{\frac{2\Lambda}{b-1}}r}{1+\sqrt{\frac{2\Lambda}{b-1}}r}\right),
\end{equation}
\begin{equation}
\frac{y+z}{y-z}=2kbe^{2b\sqrt{\frac{2\Lambda}{b-1}}k\bar{t}_{0}},
\end{equation}
with
\begin{equation}
J(\Lambda,M,r)=\frac{1}{2b}\exp\left[2\sqrt{\frac{b+1}{b-1}}\arctan\left(\sqrt{\frac{2\Lambda}{b+1}}r\right)\right].
\end{equation}

The singularity at $r=0$ would lie on the surfaces
$y^2-z^2=\frac{k}{2b}$. This singularity can be avoided by defining
either a new coordinate $\zeta=iy$, or a new coordinate $\xi=iz$.
For the first choice, the metric takes the form
\begin{equation}
ds^2=kG(\Lambda,M,r)(d\zeta^2-dz^2)+r^2d\Omega_3^2,
\end{equation}
which is positive definite in the patch $k=+1$ and has the signature
- - + + in the patch $k=-1$. The radial coordinate $r$ is then defined
as
\begin{equation}
\zeta^2+z^2=kJ(\Lambda,M,r)\left(\frac{r-\sqrt{\frac{b-1}{2\lambda}}}{r+\sqrt{\frac{b-1}{2\lambda}}}\right).
\end{equation}

On the section on which $\zeta$ and $z$ are both real,
$\sqrt{\frac{M}{\sqrt{\Lambda}}}\geq r\geq r_h\equiv\sqrt{\frac{b-1}{2\Lambda}}$
for $k=+1$, and
$\sqrt{\frac{a-1}{\Lambda}}\leq r\leq r_h$ for $k=-1$. Define the
imaginary time by $\bar{t}_{0}\rightarrow i\bar{\tau}_{0}$. This
continuation leaves invariant the form of the metric (5.7) and
therefore is compatible with the coordinate transformation $\zeta=iy$.
Then, from (5.5) we obtain
\begin{equation}
z-i\zeta=\sqrt{2b(z^2+\zeta^2)k}ie^{ib\sqrt{\frac{2\Lambda}{b-1}}k\bar{\tau}_{0}}.
\end{equation}
It follows that for this time continuation, $\bar{\tau}_{0}$ is periodic
with a period $2\pi k/b\sqrt{\frac{2\Lambda}{b-1}}$. On this nonsingular
euclidean section (which would correspond to the usual euclidean
section for the patch $k=+1$) $\bar{\tau}_{0}$ has the character of an
angular coordinate which would rotate, clockwise about the "axis" $r=r_h$
in patch $k=+1$, and anti-clockwise about the "axis" $r=0$ in patch
$k=-1$.

For the other choice of coordinates, $\xi=iz$, metric (5.2) takes
the form:
\begin{equation}
ds^2=-kG(\Lambda,M,r)(dy^2+d\xi^2)+r^2d\Omega_3^2,
\end{equation}
which is definite positive only in patch $k=-1$, and has signature
- - + + in patch $k=+1$. In this case, the radial coordinate is
defined by
\begin{equation}
\xi^2+y^2=kJ(\Lambda,M,r)\left(\frac{\sqrt{\frac{b-1}{2\lambda}}-r}{\sqrt{\frac{b-1}{2\lambda}}+r}\right).
\end{equation}

On the section on which $y$ and $\xi$ are both real, it follows now
$\sqrt{\frac{M}{\sqrt{\Lambda}}}\geq r\geq r_h\equiv\sqrt{\frac{b-1}{2\Lambda}}$
for $k=-1$, and
$\sqrt{\frac{a-1}{\Lambda}}\leq r\leq r_h$ for $k=+1$. In this case, we can
introduce the transformation $r=-i\rho, M=-i\mu, \Lambda=-\lambda$,
keeping $\bar{t}_{0}$ real. In order for this transformation to
be compatible with the coordinate redefinition $\xi=iz$, it should
leave metric (5.10) formally unchanged. This is accomplished by also
continuing the line element itseft, $ds=-id\sigma$, in accordance
with the introduction of the "tachyonic" mass $M=-i\mu$.
Thus, we have
\begin{equation}
y-i\xi=\sqrt{2b(y^2+\xi^2)k}e^{ib\sqrt{\frac{2\Lambda}{b-1}}k\bar{\tau}_{0}}.
\end{equation}

For this kind of time continuation it follows from (5.12) that it is now
the lorentzian time $\bar{t}_{0}$ which is periodic with period
$2\pi k/b\sqrt{\frac{2\lambda}{b-1}}$. On this new nonsingular
section $\bar{t}_{0}$ would have the character of an angular
coordinate which would rotate about the "axis" $\rho=0$,
clockwise in patch $k=+1$ and anti-clockwise in patch $k=-1$.

In the patch $k=+1$, the euclidean continuation (5.1) contains
both the continuation for time $\bar{t}_{0}=i\bar{\tau}_{0}$, where
the apparent singularity at $r=r_h$ is like the irrelevant
singularity at the origin of polar coordinates, provided that
$\bar{\tau}_{0}/\sqrt{\frac{b-1}{2b^2\Lambda}}$ is regarded as
an angular variable and is identified with a period of $2\pi$,
and a new continuation $r=-i\rho$, that also implies the "tachyonic"
continuations $M=-i\mu, \Lambda=-\lambda, ds=-id\sigma$, where
the singularity at $\rho=0$ becomes again harmless if
$\bar{t}_{0}/\sqrt{\frac{b-1}{2b^2\Lambda}}$ is regarded as an
angular variable with period $2\pi$. For the patch $k=-1$, the
continuation (5.1) leads to the same instantonic sections as for
the patch $k=+1$, but now $\bar{t}_{0}=i\bar{\tau}_{0}$ corresponds
to the section inside the horizon $r=r_h$, up to $r=r_{min}$, and
$r=-i\rho, M=-i\mu, \Lambda=-\lambda, ds=-id\sigma$ define the
section outside the horizon $\rho=\rho_h$, up to $\rho=\rho_{max}$,
with $\bar{\tau}_{0}$ and $\bar{t}_{0}$ rotating, respectively,
about $r=0$ and $\rho_{h}$, anti-clockwise in both cases.

On any ($\bar{\tau}_{0},r$) plane in the coordinate patch $k=+1$
we can define$^{18}$ the transition amplitude
$\langle\bar{\tau}_{02}\mid\bar{\tau}_{01}\rangle$ from the
surface $\bar{\tau}_{01}$ to the surface $\bar{\tau}_{02}$ as
given by the action of the first instanton. This action can be
expressed as the surface area of the circular sector limited by
times $\bar{\tau}_{01}$ and $\bar{\tau}_{02}$ of a circle centered
at $r=r_h$, with radius $r_{max}$. Similarly, on the plane
($\bar{t}_{0},\rho$) in patch $k=-1$, the transition amplitude
$\langle\bar{t}_{02}\mid\bar{t}_{01}\rangle$ from the surface
$\bar{t}_{01}$ to the surface $\bar{t}_{02}$ is dominated by the
action of the second instanton and is given by the surface area
of the sector limited by $\bar{t}_{01}$ and $\bar{t}_{02}$, from
$\rho=\rho_{min}$ to $\rho=\rho_{h}$, on a circle centered at
$\rho=0$. An asymptotic observer in the patch would interpret
these transition amplitudes as providing the probability of the
occurrence in the vacuum state of, respectively, a five-dimensional
black hole with mass $M$ and a five-dimensional white hole with
mass $\mu$. In the coordinate patch $k=-1$, the observer would
obtain the same interpretation, but for a black hole with mass
$M$ or a white hole with mass $\mu$.

\section{\bf Thermal emission and closed timelike curves}
\setcounter{equation}{0}

At the end of Sec. IV, we argued that the existence of closed
timelike curves (CTCs) in the spacetime under consideration is
classically disallowed because we cannot identify maximum curves
belonging to different coordinate patches. However, in what follows
we shall show how such CTCs can still be present in the semiclassical
treatment.

The semiclassical regime for kinked spacetimes with event horizons
can simply be achieved by considering$^{9}$ the mathematical
implications imposed by the fact that time $\bar{t}$ enters the
Kruskal coordinates $U,V$ in the form of the dimensionless
exponent $\sqrt{\frac{2\Lambda b^2}{b-1}}k\bar{t}$. Hence, the
argument of the logarithm in (4.10) becomes square rooted and,
therefore, the expression for the time $\bar{t}$ entering the
definition of coordinates $U,V$ should be generalized to$^{9}$:
\begin{equation}
\bar{t}\rightarrow\bar{t}_{g}=\bar{t}+i\pi\kappa(1-\kappa)k\sqrt{\frac{b-1}{8b^2\Lambda}},
\end{equation}
where $\bar{t}$ is given by (4.10) and $\kappa=\pm 1$.
For $\kappa=+1$, $\bar{t}_{g}=\bar{t}$, and for $\kappa=-1$, the
points ($\bar{t}-ik\pi\sqrt{\frac{b-1}{2b^2\Lambda}},r,\Omega_3$)
in each patch are actually the points in the same patch obtained by
refletion in the bifurcation point $U,V=0$, keeping the Kruskal
metric real and unchanged.

We note that one can still recover the standard kink metric (3.5)
from the general metric (4.1) if we redefine the Kruskal coordinates
as follows:
\begin{equation}
U=\tilde{U}=\pm 2b\kappa J(\Lambda,M,r)e^{k\sqrt{\frac{2b^2\Lambda}{b-1}}\bar{t}_{c}}
\left[\frac{\sqrt{\frac{b-1}{2\Lambda}}-r}{\sqrt{\frac{b-1}{2\Lambda}}+r}\right]
\end{equation}
\begin{equation}
V=\tilde{V}=\pm k\kappa e^{-k\sqrt{\frac{2b^2\Lambda}{b-1}}\bar{t}_{c}},
\end{equation}
where
\begin{equation}
\bar{t}_{c}=\bar{t}+i\pi\kappa k\sqrt{\frac{b-1}{2b^2\Lambda}},
\end{equation}
with $\bar{t}=-Re\left(\int\tan 2\alpha dr\right)$.

This choice leaves the expressions for $UV=\tilde{U}\tilde{V}, F, r$
and the Kruskal metric real and unchanged. For $\kappa=-1$, Eqns.
(4.8) and (4.9) become the sign-reversed of (6.2) and (6.3),
respectively; i.e.: the points
($\bar{t}-ik\pi\sqrt{\frac{b-1}{2b^2\Lambda}},r,\Omega_3$)
on the Kruskal diagrams of the two coordinate patches are the
points in the new regions for $r_h\leq r\leq r_{max}$ on the
same diagrams, obtained by reflecting in the origins of the
respective $U,V$ planes, preserving the Kruskal metric real
and unchanged. This leads to identification of hyperbolae in
the new regions for $r_h\leq r\leq r_{max}$ with hyperbolae in
the original regions for the same values of $r$. We note that
the existence of such identifications in turn amounts to both
the kind of periodicity required by Hawking thermal radiation$^{19}$
in each patch, and the existence of CTCs.

Using the procedure of Hartle and Hawking$^{19}$ to study the
evolution of a scalar field along null geodesics that start
at $r_{max}$ on the physical regions of the Kruskal diagrams
for $\kappa=-1$, it becomes now quite natural (in particular,
physical time need not be made complex) to obtain that an observer
in the exterior original region of the patch $k=+1(-1)$ will
measure an isotropic thermal bath of background radiation with
positive(negative) energy, at the temperature
\begin{equation}
T=\frac{2\pi}{\sqrt{\frac{2b^2\Lambda}{b-1}}}.
\end{equation}
This temperature exactly corresponds to the period of the periodic
time $\bar{\tau}_{0}$ that we obtained for the instantons in Sec. V.

On the other hand, since in the semiclassical description, we can
identify maximum surfaces of the physical regions with those of
the nonphysical regions in each coordinate patch separately, we
recover allowance for the null geodesics that start at the maximum
surface in region I$_{+}$ of patch $k=+1$ (Fig. 1)
to continue propagating on
patch $k=-1$, after the maximum surface of region III$_{-}$,
first through region II$_{-}$ and then through IV$_{-}$,
up to the minimum surface of the latter new region. Since this
surface can be identified with the similar surface in region III$_{+}$
of patch $k=+1$, the considered null geodesics can thereafter
propagate into the region IV$_{+}$ and, again by quantum identification
of maximum surfaces, come back to their starting points on the
maximum surface of region I$_{+}$, in patch $k=+1$.

Hence, null geodesics starting from large original regions can
still "loop back" to arrive to their starting points, so
completing a CTC, provided such a CTC is involved at a thermal
radiation process preventing any information to flow from or to
the CTC. Gibbons and Hawking$^{10}$ have recently suggested the
possibility of linking the presence of kinks to the presence
of CTCs. However, Chamblin and Penrose$^{20}$ have subsequently
shown that CTCs are not a classical requirement for kinked
spacetimes. Our result therefore fits in with these works,
since we obtain that CTCs are linked to kinks only when matter
traveling through these spacetimes is considered
quantum-mechanically.

\section{\bf Conclusion}

In this work, we have constructed the spherically-symmetric
spacetime metric describing the geometry of wormholes which
are asymptotically anti-de Sitter, generalizing this metric
to the case where the light cones are allowed to tip over on
hypersurfaces. By using the kink geometry, a relation
between four-dimensional wormholes and five-dimensional
Tangherlini black holes has been established in the anti-de
Sitter background.

We have seen that at least two coordinate patches are needed
for a complete description of the kink. Since the presence
of an event horizon makes the kink metric geodesically
incomplete, we have maximally extended this metric by resorting
to the Kruskal technique. On each spacelike slice of the kinked
Tangherlini anti-de Sitter space, the kinked wormhole results as
a bridge between two black holes, one in each patch, at a surface
inside the horizon. The space of this wormhole can be completely
described on the physical regions of the maximal analytical
extension. The extended kink metric is used
to study the instantons that can be associated with the kinked
Tangherlini-anti-de Sitter black holes. Euclidean continuation
of the metrical time of this kink implies both a continuation
of the time entering the Tangherlini metric and a continuation
of the radial coordinate, removing singularity at the origin.

It has been also shown that the black hole in one coordinate patch
emits a thermal bath formed by the antiparticles to the particles
that make up the radiation emitted by the hole described in the
other patch. Finally, we have also seen that, although this
spacetime kink does not imply existence of closed timelike curves
classically, when the involved black holes emit Hawking radiation,
the kink is always endowed with the formation of such curves.

\acknowledgements

\noindent This
research was supported by DGICYT under research project N§
PB94-0107, and by MEC Spanish German Joint Action N§ 161.B.
The author thanks U. Kasper for hospitality at Potsdam
University where part of this work was done.

%\pagebreak

\begin{center}
{\bf Figure Legend}
\end{center}

Fig. 1.- Kruskal diagrams for the two coordinate patches of the
one-kink anti-de Sitter wormhole spacetime. The trajectories for
classical geodesics are shown as continuous lines, and the
quantum identifications of maximum surfaces and the subsequent
geodesic trajectories as dashed lines.


\begin{references}

\bibitem {1} S.W. Hawking, in {\it Astrophysical Cosmology,
Pontificae Scientiarum Scripta Varia}, Vatican City, eds.
H.A. Bruck, G.V. Coyne and M.S. Longair, 1982.
\bibitem {2} J.B. Hartle and S.W. hawking, {\it Phys. Rev.} D28, 2960 (1983).
\bibitem {3} S.W. Hawking, {\it Nucl. Phys.} B239, 257 (1983).
\bibitem {4} S.W. Hawking, {\it Phys. Rev.} D37, 904 (1988);
S.W. Hawking and D.N. Page, {\it Phys. Rev.} D42, 2655 (1990);
S. Coleman, {\it Nucl. Phys.} B307, 867 (1988);
P.F. Gonz\'alez-D\'{\i}az, {\it Phys. Rev.} D42, 3983 (1990);
D45, 499 (1992); {\it Nucl. Phys.} B351, 767 (1991).
\bibitem {5} J.J. Halliwell and R. Laflamme, {\it Class. Quant. Grav.}
6, 1839 (1989);
A. Zhuk, {\it Phys. Rev.} D45, 1192 (1992);
P.F. Gonz\'alez-D\'{\i}az, {\it Phys. Rev.} D40, 4184 (1989).
{\it Grav. Cosm.} 2, 45 (1996).
\bibitem {6} P.F. Gonz\'alez-D\'{\i}az, {\it Elliptic and Circular Wormholes},
IMAFF-RC-02-93; gr-qc/9306031.
\bibitem {7} C. Barcel\'o, L.J. Garay, P.F. Gonz\'alez-D\'{\i}az and
G.A. Mena Marug\'an, {\it Phys. Rev.} D53, 3162 (1996).
\bibitem {8} D. Finkelstein and C.W. Misner, {\it Ann. Phys. (N.Y.)}
6, 230 (1959); D. Finkelstein, in {\it Directions in General Relativity I},
eds. B.L. Hu, M.P. Ryan Jr. and C.V. Vishveshwara (Cambridge Univ. Press,
Cambridge, UK, 1993).
\bibitem {9} P.F. Gonz\'alez-D\'{\i}az, {\it Phys. Rev.} 51D, 7144 (1995);
{\it Grav. Cosm.} 2, 621 (1996).
\bibitem {10} G.W. Gibbons and S.W. Hawking, {\it Phys. Rev. Lett.} 69, 1719 (1992).
\bibitem {11} A. Chamblin, {\it Kinks and Singularities}, DAMTP preprint R95/44.
\bibitem {12} A. Einstein and N. Rosen, {\it Phys. Rev.} 48, 73 (1935).
\bibitem {13} R.M. Wald, {\it General Relativity} (The University of
Chicago Press, Chicago, USA, 1984).
\bibitem {14} C.W. Misner, K.S. Thorne and J.A. Wheeler, {\it Gravitation}
(Freeman, New York, USA, 1973).
\bibitem {15} M. Abramowitz and I. Stegun, {\it Handbook of
Mathematical Functions} (Dover, New York, 1972).
\bibitem {16} D. Finkelstein and G. McCollum, {\it J. Math. Phys.} 16, 2250 (1975).
\bibitem {17} M.D. Kruskal, {\it Phys. Rev.} 119, 1743 (1960).
\bibitem {18} G.W. Gibbons and S.W. Hawking, {\it Phys. Rev.} D15, 2752 (1977).
\bibitem {19} J.B. Hartle and S.W. Hawking, {\it Phys. Rev.} D14, 2188 (1976).
\bibitem {20} A. Chamblin and R. Penrose, {\it Twistor Newsletter} 34, 13 (1992).
\end{references}
\end{document}